\documentclass[a4paper]{article}
\usepackage{amsmath,amsfonts,amssymb}
\usepackage{graphicx}
\usepackage{lscape,epsf}
\usepackage[bookmarks=true]{hyperref}

\usepackage{setspace}
\def\ds{\displaystyle}
\def\bea{\begin{array}{c}}
\def\ea{\end{array}}
\def\be{\begin{equation}\bea\ds}
\def\ee{\ea\end{equation}}
\def\bee{\begin{equation}\begin{array}{rcl}\ds}
\def\eee{\end{array}\end{equation}}

\def\d{{\rm d}}

\begin{document}

\thispagestyle{empty}

\begin{flushright}
ITEP-TH-41/12\\
IIP-TH-16/12
\end{flushright}

\vspace{30pt}
\begin{center}
{\Large \textbf{On the AdS/BCFT Approach to Quantum Hall Systems}}
\end{center}

\vspace{6pt}
\begin{center}
{\large{Dmitry Melnikov$^{a,b}$, Emanuele Orazi$^{a,c}$ and Pasquale Sodano$^{a,d}$ }\\}
\vspace{25pt}
\textit{\small $^a$  International Institute of Physics, Federal University of Rio Grande do Norte, \\Av. Odilon Gomes de Lima 1722, Capim Macio, Natal-RN  59078-400, Brazil}\\ \vspace{6pt}
\textit{\small $^b$  Institute for Theoretical and Experimental Physics, \\B.~Cheremushkinskaya 25, Moscow 117218, Russia}\\ \vspace{6pt}
\textit{\small $^c$  INFN - Laboratori Nazionali di Frascati, \\Via Enrico Fermi 40, I-00044 Frascati, Italy}\\ \vspace{6pt}
\textit{\small $^d$  INFN - Sezione di Perugia, \\Via A. Pascoli, I-06123 Perugia, Italy}\\ \vspace{6pt}
\end{center}

\vspace{6pt}

\begin{abstract}
In this paper we study a simple gravity model dual to a $2+1$-dimensional system with a boundary at finite charge density and temperature. In our naive $AdS/BCFT$ extension of a well known AdS/CFT system a non-zero charge density must be supported by a magnetic field. As a result, the Hall conductivity is a constant inversely proportional to the coefficients of pertinent topological terms. Since the direct conductivity vanishes, such behaviors resemble that of a quantum Hall system with Fermi energy in the gap between the Landau levels. We further analyze the properties stemming from our holographic approach to a quantum Hall system. We find that at low temperatures the thermal and electric conductivities are related through the Wiedemann-Franz law, so that every charge conductance mode carries precisely one quantum of the heat conductance. From the computation of the edge currents we learn that the naive holographic model is dual to a gapless system if tensionless RS branes are used in the AdS/BCFT construction. To reconcile this result with the expected quantum Hall behavior we conclude that gravity solutions with tensionless RS branes must be unstable, calling for a search of more general solutions. We briefly discuss the expected features of more realistic holographic setups.
 \end{abstract}

\newpage

\section{Introduction}

In an elegant paper~\cite{Takayanagi:2011zk} Takayanagi has recently proposed an extension of the AdS/CFT conjecture~\cite{AdS/CFT} to deal with the situation, when the CFT is defined on a space with a boundary. These so called boundary CFT's, or BCFT's, naturally appear in various applications to condensed matter phenomena. The main idea behind the Takayanagi's construction is to appropriately extend the boundary of the CFT inside the bulk of the $AdS$ space in order to cut out a
physically relevant piece.

The AdS/BCFT conjecture finds its natural roots in the holographic derivation of entanglement entropy~\cite{Ryu:2006bv} and in the Randall-Sundrum model~\cite{Randall:1999vf}. Indeed, the extension of the CFT's boundary inside the bulk of the $AdS$-space may be regarded as a modification of a ``thin'' Randall-Sundrum brane, intersecting the $AdS$ boundary. For this brane to be a dynamical object one must impose Neumann boundary conditions on the metric; then, the discontinuity in the bulk extrinsic curvature across the ``defect'' will be compensated by the brane's tension. Thus, Neumann boundary conditions constitute the dynamical principle defining the additional boundaries introduced by Takayanagi. In the following we will often refer to these boundaries as to Randall-Sundrum (RS) branes.

From the point of view of the AdS/CFT applications it is interesting to see how the addition of the boundaries affects physical behavior. In particular, it would be interesting to construct a solution describing a physical system at finite temperature and charge density. The most common playground is provided by the charged $AdS_4$ black hole since this background has been already shown to encode many interesting
condensed-matter-like phenomena such as superconductivity/superfluidity~\cite{superconductivity} and strange metallic behavior~\cite{StrangeMetals}. Here, we shall analyze the charged $AdS_4$ black hole in the presence of boundaries.

A model of gauge fields in the $AdS_4$ background with boundary RS branes has been already considered by Fujita, Kaminski and Karch in~\cite{Fujita:2012fp}, who showed that the additional boundary conditions impose relevant constraints on the gauge field parameters and that, as a consequence of those constraints, there is the possibility of quantum-Hall-like behavior of the conductivity in the dual field theory. However, their approach~\cite{Fujita:2012fp} does not account for the backreaction of the gauge fields on the geometry, restricting to the geometry of the empty $AdS$ space. A natural generalization of their work would be to extend their analysis to the case of a charged black hole. In this paper we make a first step in this direction.

The physical system analyzed in this paper is based on the model proposed by Fujita \emph{et al.} Indeed, we start from the same Lagrangian for an Einstein-Maxwell system describing the gravity dual of a field theory on a half-plane. We find that, for the simple plane-symmetric black hole ansatz, only tensionless RS branes are allowed and that the background solution does not allow to model the situation with external electric fields as in~\cite{Fujita:2012fp}. Furthermore, as a result of the Neumann boundary conditions for the gauge fields, we find that the charge density $\rho$ in the dual field theory must be supported by an external magnetic field $B$. The ratio $\rho/B$, which is equal to the Hall conductivity, is a constant inversely proportional to the sum of the coefficients of the topological terms present in the gravity action: namely, a $\Theta$-term in the bulk action and a Chern-Simons term in the boundary action on the RS branes. In addition, we get zero longitudinal conductivity. All these behaviors are expected for a quantum Hall system tuned to a quantized value of the conductivity.

In the classical Hall effect $\rho$ and $B$ are independent quantities, \emph{e.g.} their ratio is a function of the density of conductance electrons $\rho$. In the quantum Hall Effect (QHE) the transverse conductivity $\sigma_H$, exhibits plateaus independent of either $\rho$ or $B$. The plateaus are generally attributed to disorder~\cite{Laughlin:1981jd}, which is responsible for localization (\emph{i.e.} existence of localized electron states). The localized states fill the gaps between the Landau levels, but do not participate in the Hall conductivity. In other words, even if one varies the density of conductance electrons, so that the Fermi energy shifts in the gap between the Landau levels, the conductivity does not change. For the same reason the longitudinal components of the conductivity vanish.

An important aspect of the QHE is the quantization of the conductivity plateaus in integer (integer QHE) or fractional (fractional QHE) numbers of flux quanta $h^2/e$. The mechanisms of the integer and fractional quantization are, however, different: while the fractional
quantization is due to the interaction between electrons~\cite{FQHE}, the integer QHE (IQHE) exists even for a non-interacting electron gas with disorder and is a consequence of gauge invariance~\cite{Laughlin:1981jd}. In the widely accepted composite fermion model the fractional QHE (FQHE) may be regarded as an IQHE for composite quasiparticles of fractional charge~\cite{CompositeModel}.

It is by now well understood that the FQHE is effectively described by Chern-Simons gauge theories~\cite{Girvin}. Furthermore, as explicitly seen in computations of the gauge effective action of planar electrodynamics~\cite{Seso}, the Hall relation $\rho=\sigma_H B$ can be derived from the Chern-Simons Lagrangian with $\sigma_H$ being a constant inversely proportional to the Chern-Simons level $k$. If, for some reason, $k$ is constrained to take integer values (for details, see~\cite{Polychronakos:1986me}), the conductivity would be also quantized.

The general philosophy of the AdS/BCFT correspondence leads naturally to a generalization of the Chern-Simons description of the QHE. Indeed, the Neumann boundary condition for the gauge fields, derived by Fujita \emph{et. al.}, turns out to be a covariant ``holographic'' form of the Hall relation. Furthermore, in the AdS/BCFT formulation, the Chern-Simons theory lives on the RS branes cutting the bulk of the $AdS$ space and the holographic Hall relation potentially leads to a more interesting dependence of $\sigma_H$ on various topological coefficients~\cite{Fujita:2012fp}.

An interesting question is to ascertain whether and how models using the AdS/BCFT correspondence may be made consistent with the physics underlying the integer or the fractional QHE. Since the classical gravity limits of the holographic correspondence enable to describe strongly coupled gauge theories one may naively expect that the AdS/BCFT correspondence may provide us with an appropriate description of the FQHE. This naive expectation is supported by the study in~\cite{StrangeMetals}, which revealed a non-Fermi liquid behavior of the charged $AdS_4$ black hole. In addition, the gravity solution describes a ``clean'' system, where disorder has still to be introduced. In this paper we provide a further analysis of the physical consequences stemming from the $AdS$ black hole solution with the aim of elucidating its relationship with realistic quantum Hall systems.

One way to test how fractional is the QHE is to look at the behavior of the heat conductivities. For a non-interacting electron gas the heat conductivity is related to the electric conductivity through the Wiedemann-Franz law. For interacting electrons, \emph{i.e.} for the FQHE, the Wiedemann-Franz law is violated~\cite{Kane} and the ratio of the transverse heat and electric conductivities becomes a non-trivial function of the filling fraction, which may yield zero, or negative, heat conductivities. Our analysis of the low temperature behavior of the transverse thermal conductivities shows that the $AdS_4$ black hole is perfectly consistent with the Wiedemann-Franz law. In particular, the $O(T)$ coefficient of the ratio, the Lorenz number, is precisely $\pi^2/3$; a well known result for non-interacting electrons.

Another characteristic feature of quantum Hall systems is the emergence of current-carrying edge states, extending along the perimeter of the sample~\cite{Halperin:1981ug}. Their existence is due to the fact that, at the edges, there is a confining potential, preventing the electrons from leaving the sample and locally raising the energy of the Landau levels. As a result, while the system remains gapped (incompressible) far from the edges, there is no gap at the edges. The edge states may then be regarded as the massless modes arising at the intersection of the Fermi energy with the filled Landau levels. Below, we show that the edge currents do not appear for a black hole solution with tensionless RS branes. In addition, we argue that there is an edge current whenever the tension of the RS branes is different from zero. This observation is consistent with the fact that a non-vanishing tension should represent a system with a geometry-induced gap in the ``bulk'' of the sample. Our analysis hints to the fact that a black hole solution with tensionless RS branes may only describe the situation where the difference between the Fermi energy and the nearest Landau level is smaller than the temperature and, thus, it is expected to be rather unstable against thermal fluctuations.

Although the behavior found for the conductivities and the absence of a gap seem to be in contrast with our expectations for a quantum Hall system, we believe that this only hints to the instability of the gravity solution with tensionless RS branes. We feel that a more appropriate solution would be the one with non-vanishing tension of the RS branes; unfortunately, an explicit form of such a solution is still lacking. We expect that this solution will modify the result for conductivities and may lead to a violation of the Wiedemann-Franz law; a preliminary analysis indicates, indeed, that RS branes with tension enable to describe situations where the electric conductivity is quantized and the tension is related to the Fermi energy, rather than being an independent parameter. Alternatively, it is interesting to
understand how disorder may be introduced in a AdS/BCFT model describing the IQHE.

There already exists an extensive literature on the application of holographic correspondence to quantum Hall physics. Here we will take advantage of the computation of the transport coefficients carried by Hartnoll and Kovtun for the dyonic $AdS_4$ black hole~\cite{Hartnoll:2007ai} and further elaborated in~\cite{Hartnoll:2007ih}. We also stress that the AdS/BCFT model analyzed in this paper is reminiscent of some recent top-down stringy constructions~\cite{Davis:2008nv}. For more details we refer the reader to these papers.

The presentation of our results is organized as follows. In section~\ref{sec:setup} we review the AdS/BCFT construction in the particular setup studied in~\cite{Fujita:2012fp}. In section~\ref{sec:model} we investigate the black hole solution constrained by the additional boundary conditions required by the AdS/BCFT construction. Section~\ref{sec:conductivities} is devoted to the discussion of the electric and thermal conductivities. The computation of the edge currents is the main topic of section~\ref{sec:edgecurrents}. Finally, in section~\ref{sec:conclusions}, we provide a critical summary of our results.


\section{Setup of the AdS/BCFT holographic model}
\label{sec:setup}

We aim at studying a gravity dual of a $2+1$-dimensional system with finite temperature and charge density living on a space $M$ with a boundary which, for the sake of simplicity, will be taken in the following sections as a half plane or a strip. Since systems at finite temperature and charge density may be dually described by charged black holes, we can describe the bulk using the abelian Einstein-Maxwell action. Namely, one may start with
\be
\label{BulkAction}
S_{N} = \frac{1}{2\kappa}\int_N\d^4 x\, \sqrt{-g}\left(R-2\Lambda\right) - \,\frac{1}{4e^2}\int_N\d^4 x\,\sqrt{-g}F_{\mu\nu}F^{\mu\nu}+\frac{\Theta}{8\pi^2}\int_N F\wedge F \,.
\ee
Following~\cite{Takayanagi:2011zk} we denote with $N$ the bulk space, with $g$ the determinant of the bulk metric $g_{\mu\nu}$ and with $\Lambda=-3/L^2$ the cosmological constant. In the following we set the gravitational constant $\kappa=2L^2$ and the $U(1)$ coupling $e=1$; the only role of $e$ will be then to define the probe limit $e\to\infty$, in which the backreaction of the matter fields on the gravity solution can be ignored. Since the last term in the bulk action is topological, it does not affect the equations of motion; however, it is relevant for the analysis of the boundary conditions. The bulk equations of motion may then be rewritten as
\be
\label{einstein}
R_{\mu\nu} - \frac{R}{2}\,g_{\mu\nu} - \frac{3}{L^2}\,g_{\mu\nu} = \kappa\,F_{\mu\rho}{F_{\nu}}^{\rho} - \frac{\kappa}{4}\,g_{\mu\nu}F_{\rho\sigma}F^{\rho\sigma}\,,
\ee
\be
\label{maxwell}
\nabla^\mu F_{\mu\nu} = 0\,.
\ee

The dual theory lives on a $2+1$ dimensional space $M$ with a boundary which, following~\cite{Nozaki:2012qd}, we denote with $P$. For a proper definition of the holographic model it is natural to require that this $1+1$ dimensional boundary is extended to the bulk to make a $2+1$ dimensional hypersurface $Q$~\cite{Takayanagi:2011zk}, so that $\partial N = Q\cup M$ and $\partial Q = \partial M = P$. According to Takayanagi, for this hypersurface $Q$ to be dynamical, one needs to introduce the following boundary action on $Q$
\be
\label{QAction}
S_Q = \frac{1}{\kappa}\int_Q \d^3 x\,\sqrt{-h}\left(K - \Sigma \right) + S_Q[{\rm matter}]\,,
\ee
where $h$ is the determinant of the induced metric $h_{ab}$ on the boundary $Q$ and $K$ is the extrinsic curvature on $Q$. We denote by $\Sigma$ the corresponding cosmological constant, which may also be regarded as the tension of the brane $Q$. In the following we shall use $T$ only to account for the temperature of the dual field theory. As pointed out in the introduction, the boundary $Q$ plays a role analogous to the cutoff branes in the Randall-Sundrum model; in the following, we shall refer to the additional boundaries required by the AdS/BCFT construction as RS branes.

To have a well-defined variational principle on a space with boundaries one needs to specify the boundary conditions. Variation of the bulk~(\ref{BulkAction}) plus boundary~(\ref{QAction}) action in response to a variation of the metric yields the following boundary term
\be
\label{gVariation}
\delta(S_N+S_Q)[\delta g_{\mu\nu}] = \frac{1}{2\kappa}\int_Q \d^3 x\, \sqrt{-h}\left(K_{ab}-(K-\Sigma)h_{ab}\right)\delta h^{ab} - \frac{1}{2}\int_Q \d^3 x\,\sqrt{-h}\, T_{ab}\,\delta h^{ab}\,,
\ee
where $T_{ab}$ is the boundary stress-energy tensor:
\be
T_{ab} = -\frac{2}{\sqrt{-h}}\,\frac{\delta }{\delta h^{ab}}\,S_Q[{\rm matter}]\,.
\ee
For the variation to vanish one could impose a Dirichlet boundary condition on the metric $\delta h^{ab}=0$. However, if $Q$ has to be determined dynamically, one should employ Neumann boundary conditions. Namely, one should require that
\be
\label{Neumann-metric}
K_{ab}-(K-\Sigma)h_{ab}= T_{ab}\,.
\ee
As a result the shape of $Q$ and the induced metric arise as a solution of dynamical equation~(\ref{Neumann-metric}), rather than as an explicit boundary condition.\footnote{This is also a natural boundary condition from the point of view of the string theory orientifold construction considered in~\cite{Fujita:2011fp}.}

To study quantum Hall systems we chose the boundary matter action as in~\cite{Fujita:2012fp}:
\be
S_Q[{\rm matter}]= \frac{k}{4\pi}\int_Q A\wedge F \,.
\ee
Notice that, for this choice of the action, $T_{ab}=0$ and, as in equation~(\ref{gVariation}), the variation of the gauge fields induces a boundary term
\be
\label{AVariation}
\delta (S_N+S_Q)[\delta A_{\mu}] = 2\int_Q \left(\frac12\,\ast F +\left(\frac{\Theta}{8\pi^2}+\frac{k}{4\pi}\right)F\right)\wedge\delta A\,; \ee
the Neumann boundary conditions then imply that
\be
\label{Neumann-gauge0}
c_1\,*F + c_2\,F\big|_Q=0\,,
\ee
where, for simplicity, we have set
\be
\label{TopologicalCoeffs}
c_1= \frac12\,, \qquad c_2 = \frac{\Theta}{8\pi^2}+\frac{k}{4\pi}\,.
\ee

In the next section we shall analyze the solutions to the equations of motion~(\ref{einstein}) and~(\ref{maxwell}) constrained by the Neumann boundary conditions~(\ref{Neumann-metric}) and~(\ref{Neumann-gauge0}). Here, we would like to complete the analysis of the gravity action needed by the AdS/BCFT construction. Namely, the variation of the action gets contributions also from the boundary $M$ where, in accordance with the general lore of the AdS/CFT correspondence, one must impose Dirichlet boundary conditions. Following the standard procedure~\cite{Henningson:1998gx} we introduce the following counter terms to render the action finite,
\be
\label{counterterms}
S_{M} = \frac{1}{\kappa}\int_M\d^3x\,\sqrt{-\gamma}\left(K-\frac{1}{L}\right).
\ee
Here $\gamma$ is the determinant of the induced metric on $M$, which is fixed by the Dirichlet boundary conditions to be the Minkowski metric $z^2 \gamma_{\mu\nu}|_{z\to 0}=\eta_{\mu\nu}$ of the dual gauge theory.

Finally, one needs to introduce additional boundary terms at the boundary $P$:
\be
\label{PAction}
S_P = \frac{1}{\kappa}\int_P \d^2x\, \sqrt{-\sigma} (\theta -\pi) - \frac{k}{4\pi}\int_P\d^2x\, A_t A_x\,.
\ee
The first term was already introduced in~\cite{Nozaki:2012qd}. It accounts for the singularity of the boundary $P$. Indeed in~(\ref{gVariation}) we omitted a term, which is a total derivative on $Q$, and thus lives only on the boundary $P$. A similar term comes from the variation of the counter terms~(\ref{counterterms}) on $M$. In equation~(\ref{PAction}) $\pi-\theta$ ($0\leq \theta\leq \pi$) is the angle between the two outward pointing unit normal vectors $n_{M}$ and $n_{Q}$ at the boundary $P$ (\emph{e.g.}~\cite{Hayward:1993my}). When $\theta=\pi$ the two normal vectors coincide and the boundary becomes smooth. The second term in~(\ref{PAction}) is required by gauge invariance. To ensure that the Chern-Simons action on $Q$ is gauge invariant we choose the gauge condition as $\partial_t A_x=0$. With this choice the boundary term protects the residual gauge transformations. Notice that, in~(\ref{PAction}), $x$ and $t$ are generic (\emph{i.e.} not
yet specified) coordinates parameterizing $P$.

Collecting all the pieces together the total action proposed to describe the AdS/BCFT holographic model of a quantum Hall system is given by
\be
\label{TotalAction}
S = S_N + S_Q  + S_{M}+ S_P\,.
\ee


\section{The model of a half-plane with a finite charge density}
\label{sec:model}

\begin{figure}
\begin{minipage}{0.45\linewidth}
\begin{center}
\includegraphics[width=\linewidth]{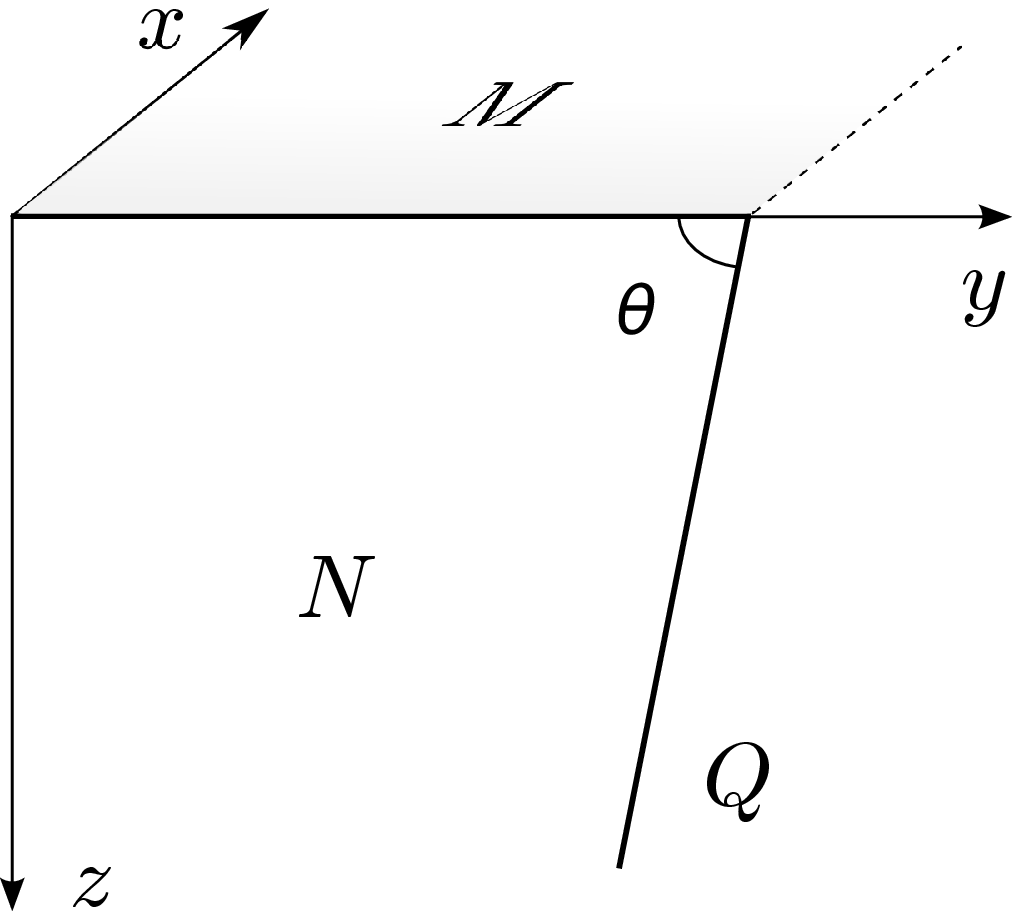}
(a)
\end{center}
\end{minipage}
\hfill{
\begin{minipage}{0.45\linewidth}
\begin{center}
\includegraphics[width=\linewidth]{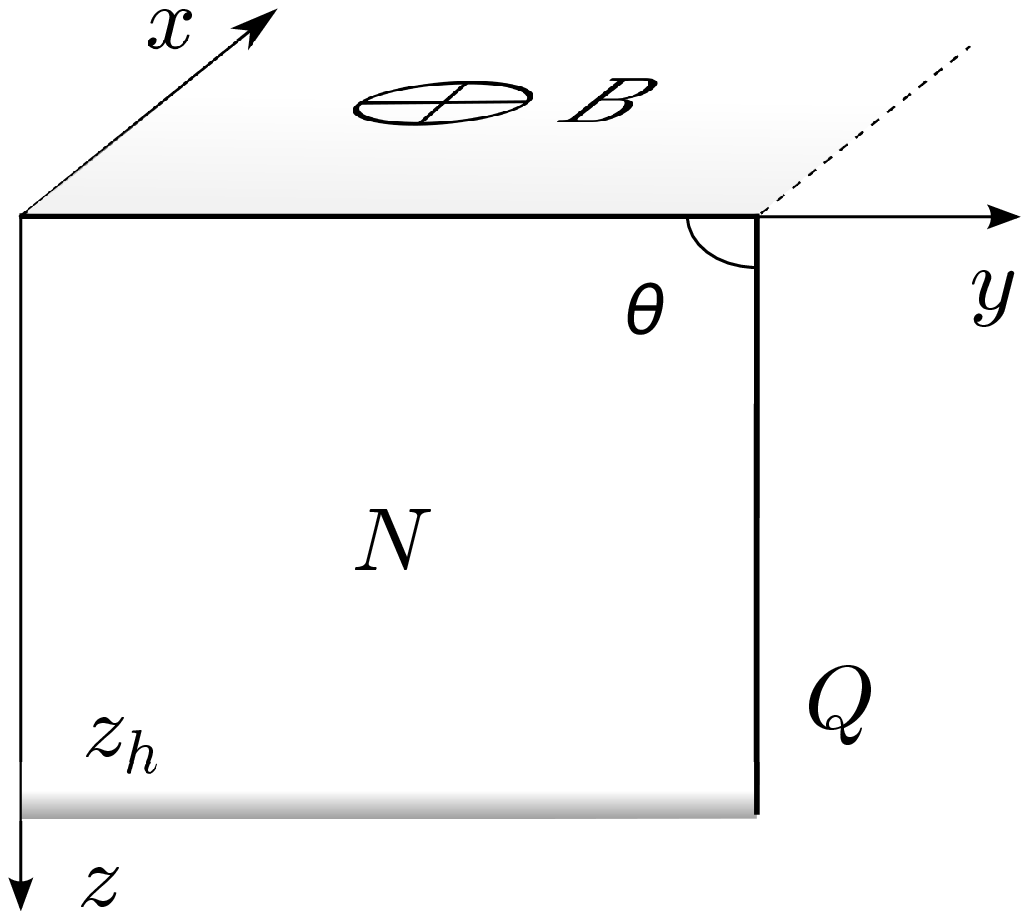}
(b)
\end{center}
\end{minipage}
} \caption{ (a) AdS/BCFT geometry of the half-plane $M$. $N$ denotes the part of the bulk space bounded by $M$, horizon of the black hole at $z_h$ and RS brane $Q$. (b) Dyonic black hole solution ($\theta=\pi/2$).}
\label{fig:geometry}
\end{figure}

In this section we describe a solution to the equations~(\ref{einstein}) and~(\ref{maxwell}), satisfying the additional boundary conditions imposed by the AdS/BCFT construction. For simplicity, we take the $2+1$-dimensional manifold $M$ to be half of an infinite plane; our subsequent analysis may, however, be readily generalized to include the case of an infinite strip. We make the following simple black hole ansatz for the metric
\be
\label{ansatz-metric}
\d s^2 = \frac{L^2}{z^2}\left(- f(z)\,\d t^2 + \d x^2 + \d y^2 + \frac{\d z^2}{f(z)} \right),
\ee
and for the gauge fields
\be
\label{ansatz-gauge}
A_t  =  \mu - \rho z\,, \qquad A_x  =  - By + cz\,, \qquad A_{y,z} =  0\,.
\ee
Here $x$, $y$ and $t$ are coordinates on $M$ such that $y<0$ (see figure~\ref{fig:geometry}).

A general solution to the equations~(\ref{einstein})-(\ref{maxwell}) with the ansatz~(\ref{ansatz-metric})-(\ref{ansatz-gauge}) is given by
\be
\label{metric-soln}
c=0, \qquad f(z) = 1 + (\rho^2+B^2)z^4 - C\,z^3\,,
\ee
where $C$ is an integration constant related to the mass of the black hole. The other parameters of the solution have the following interpretation in terms of the dual theory: $\mu$ is the chemical potential, $\rho$ is the charge density and $B$ is the external magnetic field, normal to the plane $M$. The largest of the positive roots $z_h$ of the function $f(z)$ -- the so called blackening factor -- determines the horizon of the black hole.

It is sometimes convenient to switch from a description in terms of the mass of the black hole to a description in terms of $z_h$. Upon defining
\be
Q^2=(\rho^2+B^2)z_h^4\,,
\ee
$f(z)$ takes the form
\be
f(z)=1-(1+Q^2)\left(\frac{z}{z_h}\right)^3 + Q^2\,\left(\frac{z}{z_h}\right)^4,
\ee
where both $Q$ and $z_h$ are now functions of the mass and the charge of the black hole. In terms of $Q$ and $z_h$ the black hole temperature reads
\be
T=\frac{3-Q^2}{4\pi z_h}\,.
\ee

To have a finite energy density in the dual theory one must require that $A_t(z_h)=0$; as a result, $\rho$ and $\mu$ are related as:
\be
\mu = \rho\,{z_h}\,.
\ee
The thermodynamics of this black hole was already discussed in~\cite{Hartnoll:2007ai} and~\cite{Hartnoll:2007ih}. One can describe the dual system by four thermodynamic variables: volume, temperature $T$ (or entropy $S$), charge density $\rho$ and magnetic field $B$. The entropy density is given by
\be
{s}= \frac{\pi}{z_h^2}\,.
\ee
From~\cite{Hartnoll:2007ai} the energy density $\varepsilon$ in the dual system  may be written as a function of $\rho$, $B$ and the entropy density $s$ as
\be
\label{EnergyDensity}
\varepsilon(s,\rho)= \frac{1}{2}\left(\frac{s}{\pi}\right)^{3/2}\left(1+\frac{\pi^2\rho^2}{s^2}+\frac{4\pi^2B^2}{s^2}\right)\,.
\ee

In the AdS/BCFT correspondence the parameters $\rho$, $s$ and $B$ will not be independent due to the boundary conditions at the RS brane $Q$. Let us start with the boundary condition~(\ref{Neumann-metric}). We assume that $Q$ is parameterized by the equation $y=y(z)$ and, to find the extrinsic curvature, we define an outward unit vector normal to $Q$
\be
\label{normalQ}
(n^t,n^x,n^y,n^z)=\left(0,0,\frac{z}{L\sqrt{1+f(z)y'(z)^2}}\,,-\,\frac{z f(z)y'(z)}{L\sqrt{1+f(z)y'(z)^2}}\right)\,;
\ee
then, the pullback of equation~(\ref{Neumann-metric}) to $N$ reads:
\be
\label{Neumann-metric2}
K_{\mu\nu}-(K-\Sigma)h_{\mu\nu}=0\,,
\ee
where the extrinsic curvature is given by
\be
K = h^{\mu\nu}K_{\mu\nu}\,, \qquad K_{\mu\nu}={h_{\mu}}^\rho {h_{\nu}}^\sigma\nabla_{\rho}n_\sigma\,,
\ee
and ${h_{\mu}}^\nu$ is the projector onto $Q$:
\be
{h_{\mu}}^\nu = {\delta_\mu}^\nu - n_\mu n^\nu\,.
\ee

The simplest path to solve equation~(\ref{Neumann-metric2}) is to set $\rho=B=C=0$;  this enables to recover the result
of~\cite{Takayanagi:2011zk}. The metric is pure $AdS_4$ and $Q$ is given by the solution to
\be
\label{Qtension}
y'(z)=\, \frac{L\Sigma}{\sqrt{4-L^2\Sigma^2}}\,.
\ee
Setting $L\Sigma=-2\cos\theta$ one gets
\be
\label{QprofileT}
y= - z\cot\theta\,,
\ee
\emph{i.e.} $Q$ can be described as a plane intersecting the boundary $M$ at an angle $\theta$, as shown in figure~\ref{fig:geometry}(a).

More generally, the only possibility to get an RS brane with non-zero tension from~(\ref{ansatz-metric}) is to set $f'(z)=0$; this means that no black hole with finite tension of $Q$ can be constructed from our simple ansatz for the black hole. For $\Sigma=0$, however, a black hole~(\ref{metric-soln}) solution is allowed. In the latter case the RS brane $Q$ is just perpendicular to the boundary $M$ (figure~\ref{fig:geometry}(b)):
\be
\label{Qzerotension}
Q\,:\quad y={\rm const}\,, \quad \theta=\frac{\pi}{2}\,.
\ee
The existence of non-trivial gravity solutions with non-zero tension of the RS branes was recently addressed in~\cite{Nozaki:2012qd}. It was shown that such solutions do exist; it is very interesting, indeed, to ascertain whether such solutions may be adapted to describe charged black holes.

Let us now turn to the discussion of the restrictions imposed by the boundary condition~(\ref{Neumann-gauge0}) for the gauge fields. Although we already know that, in our simple setup, the background gauge fields are only allowed for zero tension, it will be instructive to keep the discussion more general using~(\ref{QprofileT}) and setting $\theta=\pi/2$ only at the end of our derivation. In component notation, the Neumann boundary condition~(\ref{Neumann-gauge0}) may be written as
\be
\left.\frac{c_1}{2}\,\sqrt{g}\,\epsilon_{\mu\nu\rho\sigma}F^{\rho\sigma} + c_2\,F_{\mu\nu}\right|_Q=0\,.
\ee
To restrict the bulk equation on $Q$ one may project it on a 3-form orthogonal to the normal vector $n^\mu$ so that
\be
\label{Neumann-gauge}
- c_1\,\sqrt{g}\,n_\nu F^{\nu\mu} + \frac{c_2}{2}\,n_\nu\epsilon^{\nu\mu\rho\sigma} F_{\rho\sigma}=0\,.
\ee
Equation~(\ref{Neumann-gauge}) then leads to
\be
\label{Neumann-gauge2}
\begin{array}{ccc}
0 & = & c_1\rho\cos\theta - c_2(c\sin\theta+B\cos\theta)\,,
\\ 0 & = & c_1(B\sin\theta-c\cos\theta) + c_2\rho\sin\theta\,.
\end{array}
\ee
For $c=0$ and tensionless $Q$, $\theta=\pi/2$, one gets
\be
\label{rhoBratio} \frac{\rho}{B} = -\,\frac{c_1}{c_2}\,.
\ee
We see that the density and the magnetic field are no longer two independent parameters. Since their ratio is just the Hall conductivity this is very reminiscent of the quantum Hall effect, where this ratio is independent of both $\rho$ and $B$ and is inversely proportional to the topological coefficients. Notice however that the $\theta$ dependence of equation~(\ref{Neumann-gauge2}) shows that, in the limit of maximal tension of the RS brane, ($\theta=0$), the ratio $\rho/B$ becomes proportional to $c_2/c_1$.

These results are particular cases of a more general one obtained by Fujita \emph{et al.} But unlike in~\cite{Fujita:2012fp}, here we consider a full backreacted black hole solution, which further reduces the freedom in the choice of the parameters. In particular, the black hole ansatz~(\ref{ansatz-metric}) does not allow to include additional background fields such as the electric field $E_y$ in~\cite{Fujita:2012fp}. Thus, it would be interesting to look for generalizations of the simple gravity solution analyzed in this paper.

Equations~(\ref{Neumann-gauge2}) are invariant under the shift $\theta\to\pi+\theta$, which makes explicit the invariance of~(\ref{Neumann-gauge}) under the change of sign of $n^\mu$. As a result, the same solution may live on an infinite strip of finite width $-l<y<0$. In fact, since the background fields (the charge density and the magnetic field) do not depend on the spatial coordinates, \emph{i.e.} they do not fall off at infinity, one must worry about boundary conditions at the boundary $l\to \infty$ even when dealing with a model living on the half-infinite plane. For an infinite plane one can always assume periodicity. Thus, $M$ is always topologically equivalent to a compact space; as a result it is more appropriate to picture the spatial part of $M$ as being a closed strip rather than an infinite plane.


\section{Electric and thermal conductivities}
\label{sec:conductivities}

To further test the AdS/BCFT construction against the behavior of a realistic quantum Hall system we compute, in this section, the transport properties emerging from the AdS/BCFT construction: for this purpose, one has to analyze the linear response of the system to a small perturbation of the external sources. In particular, one has to compute the electric and heat conductivities as a response to a small applied electric field or to a temperature gradient. One cannot consider the two kinds of conductivities separately, since the dual fluctuations of the gravity fields couple them to each other.

The transport properties of the dyonic black hole were originally studied in~\cite{Hartnoll:2007ai}. In our AdS/BCFT model the action used in~\cite{Hartnoll:2007ai} is modified by various boundary terms and one can, in principle, expect that the boundary terms may lead to a modification of previously known holographic results for conductivities. In the following we shall show that, for a gravity solution with tensionless RS branes, the Kubo formulae for electrical and thermal conductivities do not get contributions from these additional boundary terms. The derivation goes as follows.

The transport coefficients are encoded in the 2-point correlation functions. The holographic prescription~\cite{GKPW} relates the 2-point functions to the second derivatives of the total renormalized on-shell action~(\ref{TotalAction}) with respect to the sources -- the $z\to 0$ asymptotic values of the gravity fields. Since, in the linear response theory, only linearized equations are relevant one should consider only the quadratic part of the action. To compute the charge transport one has, then, to turn on small electric fields $\delta A_x$ and $\delta A_y$ on the boundary and derive the linearized equations for the corresponding fluctuations of the bulk gauge fields; at finite temperature, the electric currents will be accompanied by the heat currents and one needs to turn on the bulk fluctuations of the metric as well:
\be
\label{variations}
\delta A_{x,y} =   A_{x,y}(z) e^{-i\omega t}\,, \qquad  \delta g_{tx,ty}  =   g_{tx,ty}(z) e^{-i\omega t}\,.
\ee
The linearization of the system of equations~(\ref{einstein})-(\ref{maxwell}) with respect to the fluctuations over the black hole solution may be reduced to the following pair of complex equations written in terms of complex fluctuations $A=A_x+iA_y$ and $G=g_{tx}+ig_{ty}$:
\be
\label{complex1}
A'' + \frac{f'}{f}\,A' + \frac{\omega^2}{f^2}\, A - \frac{\rho z^2}{fL^2}\,G' - 2\,\frac{\rho z}{fL^2}\,G + \omega\,\frac{ B z^2}{f^2L^2}\,G \,= \, 0 \,,
\ee
\be
\label{complex2}
\omega \left(G'+\frac{2}{z}\,G - 4L^2\rho A \right) - \, 4{z^2}\rho B G + 4L^2B f A' \, = \, 0\,,
\ee
where $f$ is the blackening factor given in equation~(\ref{metric-soln}). The general solutions for the metric and the gauge field fluctuations in the background~(\ref{ansatz-metric})-(\ref{metric-soln}) take the following asymptotic form at the boundary $z\to 0$. Namely, one has that
\be
\label{gboundary}
\delta g_{tj}= \frac{g_{tj}^{(0)}}{z^2} + g_{tj}^{(1)} z+ O(z^2)\,,
\ee
\be
\label{Aboundary}
\delta A_{j}= A_{j}^{(0)} + A_{j}^{(1)} z  + O(z^2)\,.
\ee
Holography then prescribes to identify the coefficients of the leading terms with the sources in the dual field theory, that is with the electric field and the temperature gradient,
\be
A_{j}^{(0)}= -\frac{E_{j}}{i\omega}\,, \qquad g_{tj}^{(0)}= - \frac{\nabla_{j}T}{i\omega T}\,,
\ee
while the subleading coefficients should be proportional to the expectation values of the dual operators, \emph{i.e.} to the charge and to the momentum density currents $J_i$ and $T_{ti}$, respectively. Notice that the second derivative of the metric fluctuation is absent from the equations~(\ref{complex1}),~(\ref{complex2}). Therefore, $g_{tj}^{(1)}$ may be written in terms of $A_j^{(1)}$ and the sources as:
\be
\label{density-current}
g_{tj}^{(1)}=-\,\frac{4L^2B}{3\omega }\,A_j^{(1)}   + \frac{4B\rho }{3 \omega }\,g_{tj}^{(0)} +\frac{4}{3 }\,\rho L^2 A_{j}^{(0)}\,.
\ee

Computing the on-shell quadratic action yields
\begin{multline}
\label{QuadAction}
S_{o/s}^{(2)}=\int \d x\,\d y\,\d t~\left[\frac{3}{8L^4}\, g_{tj}^{(0)}g_{tj}^{(1)}- \frac{\left(1+Q^2\right)}{4z_H^3 L^4}\,g_{tj}^{(0)}g_{tj}^{(0)}-\frac{\rho}{2L^2}\,g_{tj}^{(0)} A_j^{(0)}+\frac{1}{2}\,A_j^{(0)}A_j^{(1)}\right] =
\\ = \int \d x\,\d y\,\d t~\left[-\frac{B}{2\omega L^2}\, g_{tj}^{(0)}A_j^{(1)}- \frac{\left(1+Q^2\right)\omega - 2 z_h^3B\rho}{4\omega z_h^3L^4}\,g_{tj}^{(0)}g_{tj}^{(0)}+\frac{1}{2}\,A_j^{(0)}A_j^{(1)}\right].
\end{multline}
The boundary terms in the action~(\ref{TotalAction}) do not give any finite contribution to equation~(\ref{QuadAction}) for the gravity solution with tensionless RS branes. As a result the quadratic action is the same as the one considered by Hartnoll and Kovtun~\cite{Hartnoll:2007ai}, who computed the retarded Green's functions needed for the evaluation of $J_{i}$ and $T_{ti}$  and derived the electric conductivity. The thermal conductivity was later derived by Hartnoll \emph{et al.} in~\cite{Hartnoll:2007ih}. In the following we simply adapt their results to our AdS/BCFT model.

The electric conductivity is found from the retarded Green's function $G^{R}_{J_jJ_k}(\omega)$ of 2 electric currents, through the Kubo formula
\be
\sigma_{jk} = -\lim\limits_{\omega\to 0}\frac{{\rm Im\,}G^{R}_{J_jJ_k}(\omega)}{\omega}
\ee
From~\cite{Hartnoll:2007ai} one then finds
\be
\sigma_{xx}=\sigma_{yy}=0\,,\qquad \sigma_{xy}=-\sigma_{yx}= \frac{\rho}{B}= - \,\frac{c_1}{c_2}\,.
\ee
That is, the Hall conductivity is inversely proportional to the sum of the coefficients in front of the topological terms appearing in the AdS/BCFT gravity action.

The computation of the thermal conductivity must be done with a greater care. As we learn from~\cite{MagnetizationCurrents}, in the presence of a magnetic field, the Kubo formula must be modified in order to subtract the ``magnetization currents''. The modified Kubo formula reads
\begin{eqnarray}
\alpha_{jk}&=&-\,\frac{1}{T}\,\lim\limits_{\omega\to 0}\frac{{\rm Im\,}G^{R}_{J_jQ_k}(\omega)}{\omega} + \frac{\mathcal{M}}{T}\,\epsilon_{jk}\,, \qquad \text{for thermo-electric conductivity,}\\
\kappa_{jk}&=&-\,\frac{1}{T}\,\lim\limits_{\omega\to 0}\frac{{\rm Im\,}G^{R}_{Q_jQ_k}(\omega)}{\omega} + 2 \,\frac{(\mathcal{M}^E-\mathcal{M})}{T}\,\epsilon_{jk}\,, \qquad \text{for heat conductivity,}
\end{eqnarray}
where $G^{R}_{J_jQ_k}(\omega)$ and $G^{R}_{Q_jQ_k}(\omega)$ are retarded Green's functions of the electric and heat $Q_j=T_{tj}-\mu J_j$ currents. The magnetization density $\mathcal{M}$ and energy magnetization density $\mathcal{M}^E$ were computed in~\cite{Hartnoll:2007ih} as:
\be
\mathcal{M}= -Bz_h\,, \qquad \mathcal{M}^E = \frac{\mu\mathcal{M}}{2}\,.
\ee

A low temperature expansion of the result obtained in~\cite{Hartnoll:2007ih} for $\alpha_{jk}$ and $\kappa_{jk}$ yields the thermoelectric conductivity
\be
\alpha_{xx}=\alpha_{yy}=0\,,\qquad \alpha_{xy}=-\alpha_{yx}= \frac{s}{B}= \frac{\pi}{\sqrt{3}}\,\sqrt{1+\sigma_H^2}+O(T)\,,
\ee
where we used $\sigma_H=\sigma_{xy}=\rho/B$. For the longitudinal thermal conductivity one gets
\be
\kappa_{xx}=\kappa_{yy}= \frac{s^2T}{\rho^2+B^2}=\frac{\pi^2}{3}\,T+O(T^2)\,.
\ee
As expected, the conductivity is linear with temperature. The transverse conductivity behaves as
\be
\kappa_{xy}=-\kappa_{yx}= \frac{\rho s^2T}{B(\rho^2+B^2)}=\frac{\pi^2}{3}\,\sigma_HT+O(T^2)\,.
\ee
The last equation shows that the Wiedemann-Franz law is satisfied for the transverse conductivities. In fact,
\be
\label{WFlaw}
\frac{\kappa_H}{\sigma_H}=\mathcal{L}T\,,  \qquad \mathcal{L}=\frac{\pi^2}{3}\,\left(\frac{k_B}{e}\right)^2.
\ee
Remarkably, at low temperature, the $AdS_4$ black hole yields for the Lorenz number $\mathcal{L}$ the well known result valid for non-interacting electrons. It would be interesting to see whether this has anything to do with the $AdS_2\times R^2$ near-horizon structure of the extremal $AdS_4$ black hole; indeed, the latter feature is believed to underly the non-Fermi-liquid-like behavior found in~\cite{Faulkner:2009wj}.


\section{Edge modes and geometric gap}
\label{sec:edgecurrents}

A quantum Hall system is characterized by the presence of edge currents~\cite{Halperin:1981ug}. In this section we determine the conditions under which the existence of edge currents may be accounted for by the AdS/BCFT construction analyzed in this paper. Our analysis shows that no edge current can emerge from geometries with tensionless RS branes since they may be associated only to a system with no gap in the bulk. We then argue that, for geometries with tensionful RS branes, one may induce a gap in the bulk of the sample by relating the brane tension to the Fermi energy and to the scale of the closest approach of the brane $Q$ to the horizon; in this setting edge currents may emerge.

To see how edge currents may be accounted for in the AdS/BCFT construction developed in the previous sections we look at the gauge part of the action on the RS brane $Q$:
\be
S_k = \frac{k}{4\pi}\int_Q A\wedge F - \frac{k}{4\pi}\int_P \d^2 x\, A_x A_t\,.
\ee
The $x$ component of the current density can be found as the variation of the on-shell action with respect to the asymptotic value of the gauge field component $A_x$,
\be
J^x(x^i)= \frac{\delta S_k}{\delta A_x(x^i)}\,.
\ee
On shell, the action $S_k$ becomes
\be
S_k = \frac{k}{2\pi}\int_Q \d^3 x\, A_x\partial_y A_t - \frac{k}{2\pi}\int_P \d^2x\, A_x A_t\,.
\ee
Choosing the background geometry~(\ref{QprofileT}) with non-zero tension ($\Sigma\neq 0$) of the RS brane yields for the current the following expression:
\be
\label{EdgeCurrent}
J^x (x^i)= \left.\frac{k}{2\pi}\,\partial_y A_t\right|_{Q}\Theta(y+z_h\cot\theta) - \left.\frac{k}{2\pi} \,A_t\right|_{P}\delta(y) = - \frac{k}{2\pi}\,\mu\left(\delta(y) -\frac{\Theta(y+z_h\cot\theta)}{z_h\cot\theta}\right).
\ee
Here $\Theta(y)$ is the step function with the support $y\in[-z_h\cot\theta,0]$. Its appearance is not surprising since one may expect that the RS brane will be cut by the horizon of the black hole as it happens for the geometry with zero tension of the RS brane analyzed in the previous sections. In the above derivation we have also used the relation between the charge density and the chemical potential.\footnote{The same current will flow on the other boundary $y=-\ell$ albeit in the opposite direction.}

\begin{figure}[htb]
\begin{minipage}{0.45\linewidth}
\begin{center}
\includegraphics[width=\linewidth]{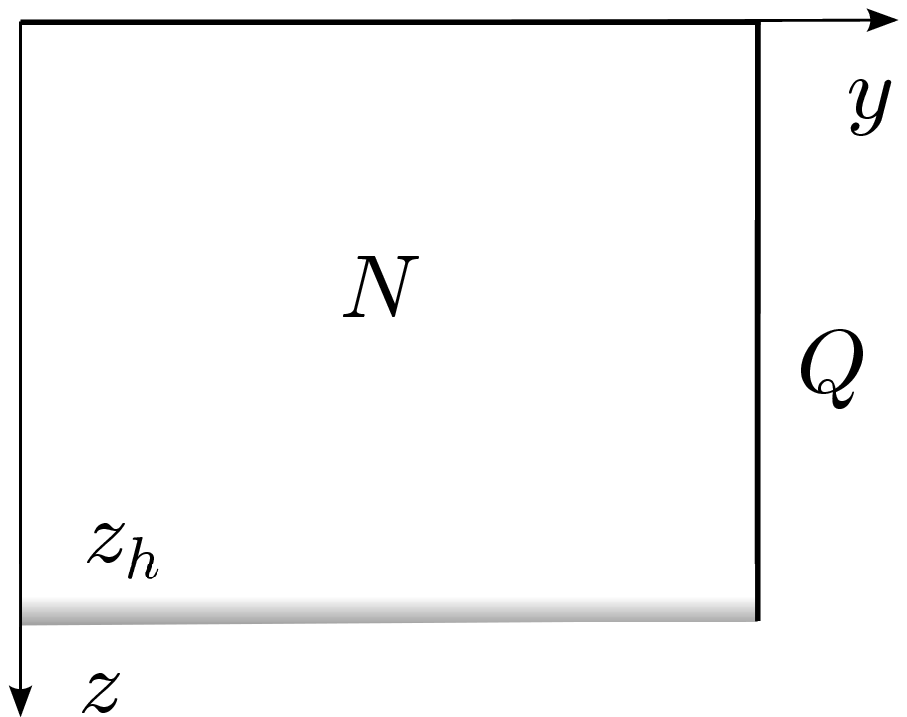}
(a)
\end{center}
\end{minipage}
\hfill{
\begin{minipage}{0.45\linewidth}
\begin{center}
\includegraphics[width=\linewidth]{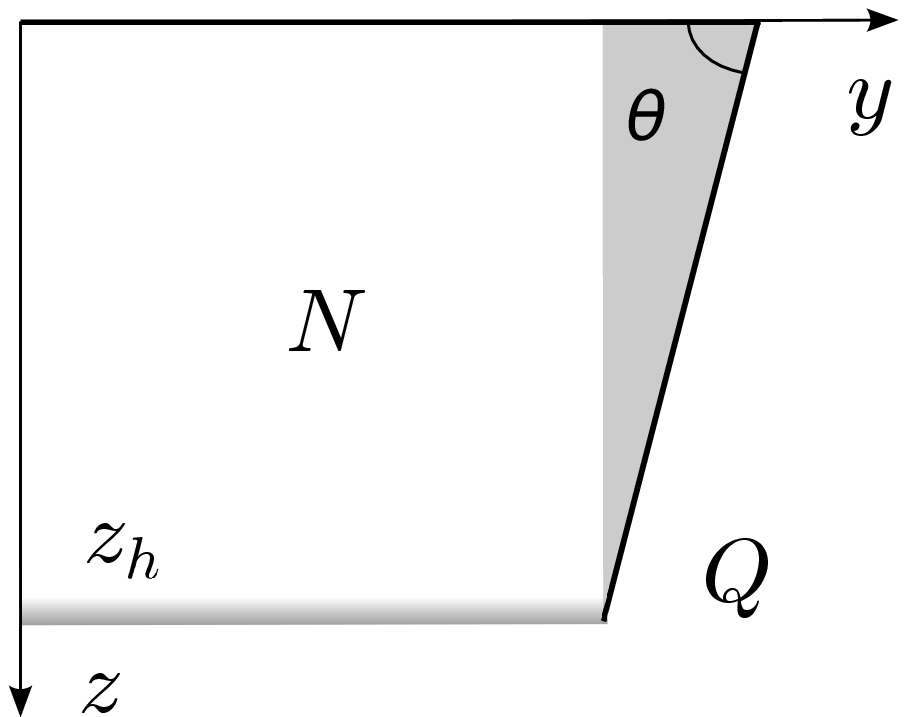}
(b)
\end{center}
\end{minipage}
}

\vspace{0.5cm}
\begin{minipage}{0.45\linewidth}
\begin{center}
\includegraphics[width=\linewidth]{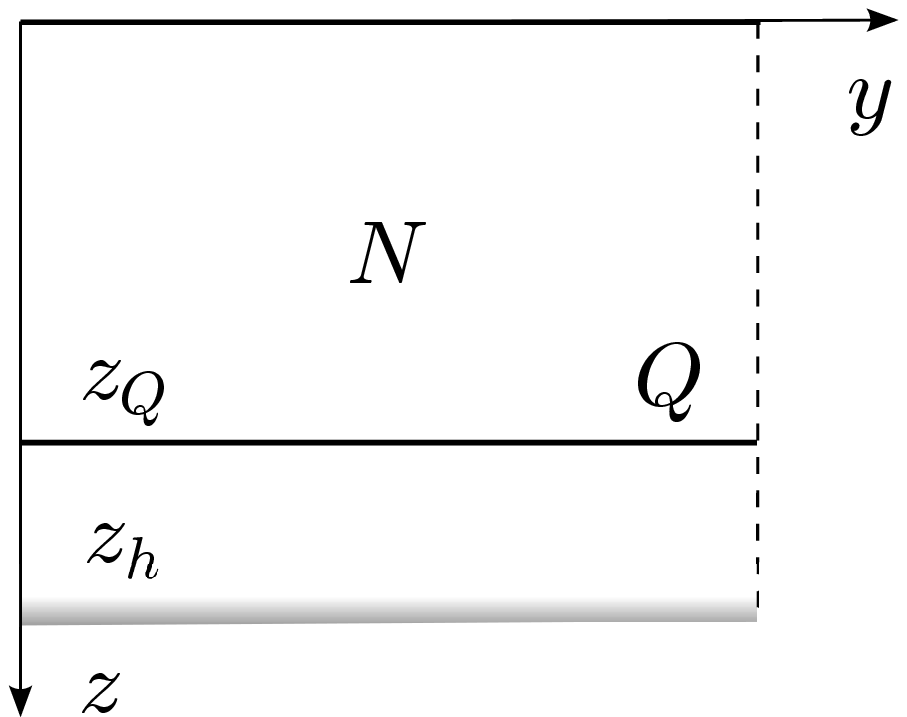}
(c)
\end{center}
\end{minipage}
\hfill{
\begin{minipage}{0.45\linewidth}
\begin{center}
\includegraphics[width=\linewidth]{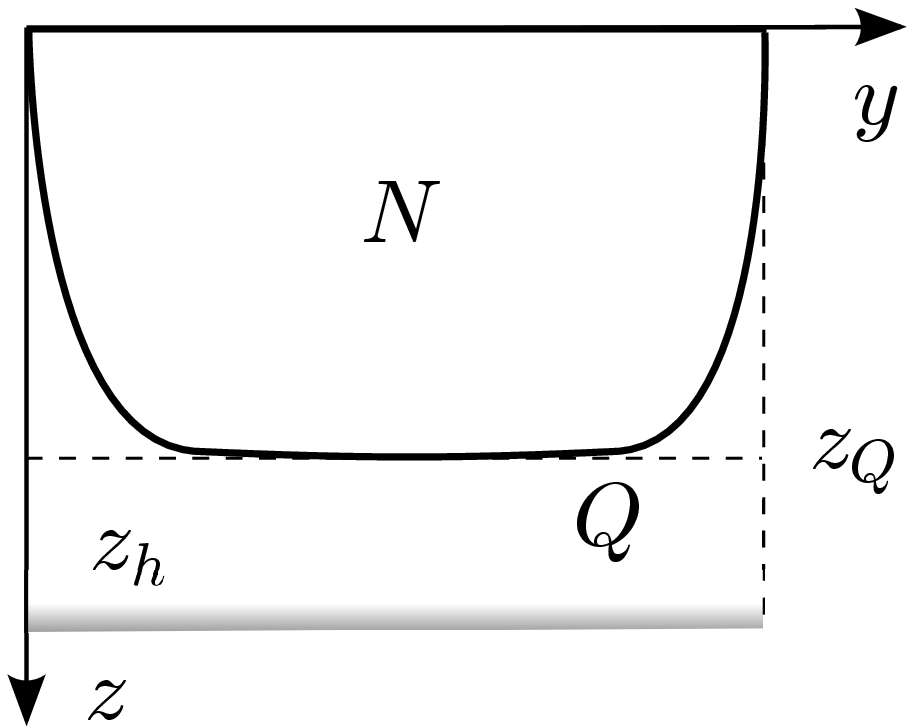}
(d)
\end{center}
\end{minipage}
}
\caption{AdS/BCFT geometry in the case of zero, ($\theta=\pi/2$)  (a), finite negative, $0<\theta<\pi/2$ (b) and minimal tension, $\theta=0$ (c). (b) and (c) can be interpreted as a position dependent and constant gap in the dual theory respectively. More realistic configurations are expected to have shape (d).}
\label{fig:gap}
\end{figure}

Let us now analyze the dependence of the current on $\theta$ or, equivalently, on the tension $\Sigma$. Although we did not find a charged black hole solution with non-zero tension, we can address its expected features in the probe limit $e\to\infty$ in~(\ref{BulkAction}), in which the backreaction of the gauge field on the metric can be neglected. When $L\Sigma\to -2$, the RS brane may be regarded as an IR cutoff, just as the original RS branes (figure~\ref{fig:gap}(c)). In this limit $\theta\to 0$ and the last term in~(\ref{EdgeCurrent}) vanishes; it is safe, then, to associate the cutoff scale with a geometry-induced gap in the dual Hall system. In a realistic quantum Hall system the gap is the difference between the Fermi energy and the energy of the closest upper Landau level -- \emph{i.e.} just the amount of energy needed to excite a charge current. Due to the potential barrier confining the electrons to the interior of the sample the Landau levels shift in the vicinity of the edge and the Fermi energy level is crossed by the lower Landau levels. ``Edge modes'' at the intersection carry then the edge current along the perimeter of the system. The edge current is given as
\be
\sigma_H=e\,\frac{\partial J}{\partial\mu}\,,
\ee
which is consistent with our previous results if one sets
\be
\sigma_H=\frac{\rho}{B}=\frac{c_2}{c_1}\,.
\ee
Indeed this is the solution to equations~(\ref{Neumann-gauge2}) with $\theta=0$. Our conventions for $c_1$ and $c_2$ yield the proper quantization of the filling fraction.

When $\theta=\pi/2$, \emph{i.e.} for the gravity solution with zero tension of the RS brane analyzed in this paper (figure~\ref{fig:gap}(a)), there is no geometric gap in the ``bulk" of the conductor, but rather an infinite gap at the edges; of course, the edge current is zero in this situation. When $0<\theta<\pi/2$, the gap is position dependent and the total edge current diffuses over a characteristic region $\Delta y = z_h\cot\theta$ (figure~\ref{fig:gap}(b)) set by the cutoff scale $z_h$.

The absence of edge currents when the RS brane is tensionless indicates that the AdS/BCFT black hole solution analyzed in this paper does not give a full description of a realistic quantum Hall system since there is no scale in the geometric description that can be associated with the energy gap needed to describe the bulk of a quantum Hall system. However, even when the RS brane is tensionless, the Hall conductivity still exhibits a plateau. The ``unphysical'' behavior the AdS/BCFT model with tensionless RS brane is likely to indicate its instability. Indeed a more general profile of the RS brane should have the shape illustrated by figure~\ref{fig:gap}(d). It would be interesting to find such tensionful solutions and analyze their stability.

For solutions of the type depicted on figure~\ref{fig:gap}(d) we expect that the tension is not a free parameter, but is dynamically related to the Fermi energy and to the scale $z_Q$ (position of the closest approach of the brane $Q$ to the horizon). The variation of the chemical potential will then correspond to the variation of $z_Q$, with $z_Q\to z_h$ as $\mu$ increases. At some value of the chemical potential one should have $z_Q=z_h$ and the gap will close. In this respect the solution in figure~\ref{fig:gap}(a) looks like a metastable branch of the profile in figure~\ref{fig:gap}(d) for $z_Q>z_h$.

Since the ``bulk'' of the sample in the configuration depicted in figure~\ref{fig:gap}(d) is gapped just as it happens for the background geometry~(\ref{QprofileT}) with non-zero tension ($\Sigma\neq 0$) of the RS brane, we expect to observe currents at the edges. It would be interesting to further understand the edge effects in such a configuration. Naively, one should expect that the conductivity is a function of $\theta$ near the edges, varying between the $\theta=0$ value ($\propto k$) and $\theta=\pi/2$ value ($\propto 1/k$). It would be interesting to ascertain whether this feature has any relationship with quantum Hall samples, where stripes of different filling fraction may appear close to the edges~\cite{Stripes}.

Transport at the edges may exhibit other very interesting phenomena. In particular, the existence of  ``upstream'' modes was conjectured in~\cite{UpstreamModes}; these modes should carry charge in the direction opposite to the classical skipping orbits typical of electrons in a magnetic field. Although no charge-carrying upstream modes has been detected experimentally, the heat transport in the upstream direction has been already observed~\cite{Venkatachalam}. This effect leads to a violation of the Wiedemann-Franz law~(\ref{WFlaw}). The violation is a result of the electron interaction: in a state with a fractional value of the filling fraction $\nu$ (FQHE) the modified Wiedemann-Franz law can be written as
\be
\label{WFlaw2}
\frac{\kappa_H}{\sigma_H}=\nu_Q\mathcal{L}T\,,
\ee
where $\nu_Q$ is a function of $\nu$, which equals unity on integer values. For fractional values of $\nu$ the function $\nu_Q(\nu)\neq 1$ and may vanish or become negative~\cite{Kane}. In the previous section we obtained $\nu_Q=1$ for a background geometry with zero tension of the RS brane; this points to the fact that this geometry may enable to describe a non-interacting system (IQHE) once disorder is included in the AdS/BCFT construction. Nevertheless, since we expect that solutions with non-vanishing tension of the RS brane will modify the computations of the conductivities (in the most naive form we can expect that the ratio in~(\ref{WFlaw}) will be $\theta$-dependent,) our AdS/BCFT approach may yield other solutions violating the Wiedmann-Franz law.


\section{Summary and concluding remarks}
\label{sec:conclusions}

We analyzed an AdS/BCFT model of a condensed matter system at finite temperature and charge density living on a $2+1$-dimensional space with a boundary. It turns out that a straightforward generalization of a known AdS/CFT solution, such as the plane-symmetric charged $AdS_4$ black hole, only allows for tensionless RS branes in the AdS/BCFT construction and requires that the static uniform charge density is supported by a magnetic field. Specifically, we found that $\rho/B$ is a constant proportional to a ratio of the coefficients appearing in the gravity action. Such a property indicates that a pertinent generalization of the $AdS_4$ black hole may describe a quantum Hall system at a plateau of the transverse conductivity.

We provided further tests of the AdS/BCFT holographic model to see how accurately it can account for the physical behaviors expected in a quantum Hall system. The AdS/BCFT construction yields that the Hall conductivity is inversely proportional to a sum of the coefficients of the topological terms appearing in the gravity Lagrangian. Namely, we got that
\be
\label{SigmaHall}
\sigma_H=\frac{\rho}{B}=-\frac{c_1}{c_2}\,, \qquad \text{where} \qquad c_2=\frac{k}{4\pi} + \frac{\Theta}{8\pi^2}\,.
\ee
Here $k$ is the level of the Chern-Simons term on the boundary and $\Theta$ is the bulk $\theta$-angle. In the QHE the conductivity is related to the number of filled Landau levels (filling fraction) by
\be
\nu=\frac{h}{e^2}\,\sigma_H= -2\pi\,\frac{c_1}{c_2}\,,
\ee
where $e^2/h$ is the magnetic flux quantum. Thus, the holographic description seems to provide results similar to the Chern-Simons description of the QHE.

As in the Chern-Simons description~\cite{Girvin,Seso}, in the AdS/BCFT model the filling is inversely proportional to the level of the Chern-Simons theory. Indeed equation~(\ref{Neumann-gauge0}) comes from the variation of the Chern-Simons Lagrangian, which lives on the RS brane $Q$. It can be seen as an extension of the covariant form of the Hall relation $\rho=\sigma_H B$. However, in the holographic AdS/BCFT model, the conductivity receives a contribution also from the bulk $\theta$-term, resulting in a renormalization of the Chern-Simons coefficient.

In the QHE, the filling fraction is quantized. It may take either integer (IQHE) or special fractional (FQHE) values. From the point of view of the Chern-Simons description, quantization of the filling fraction signifies quantization of the Chern-Simons coefficient. Unlike the non-abelian case, there is no \emph{a priori} quantization of the abelian Chern-Simons coefficient. Nevertheless, in some cases, like compact manifolds, or finite temperature, it can be argued to take integer or rational values, \emph{e.g.} see~\cite{Polychronakos:1986me}. As pointed out in section~\ref{sec:model} the  boundary conditions effectively make the manifold $M$ topologically equivalent to a compact one.  Thus, the AdS/BCFT holographic model rather easily yields the integer quantization of the Chern-Simons level $k$  which, in turn, leads to fractional values of the filling fraction and conductivity in the appropriate units.

The computation of the electrical and thermal conductivities in the AdS/BCFT model with tensionless RS branes goes along the same lines as in~\cite{Hartnoll:2007ih}. Equipped with the result of~\cite{Hartnoll:2007ih} we observed here that, at low temperatures, the coefficient of the leading $O(T)$ term yields precisely one quantum of the transverse heat conductance per quantum of the transverse electric conductance. That is the conductances satisfy the Wiedemann-Franz law for a non-interacting electron gas~(\ref{WFlaw}).

Quantum Hall systems also exhibit currents running along the edges of the sample. ln section~\ref{sec:edgecurrents} we derived an expression for the edge current, which is consistent with conventional wisdom: namely, the current is proportional to the chemical potential and to the conductivity, while its existence is associated to the existence of a gap in the ``bulk'' of the sample. Since the charged black hole solution with a tensionless RS brane analyzed in this paper cannot describe a gapped system, we find no edge current. We conclude that such a black hole does not describe completely a realistic quantum Hall system; we conjecture indeed that such black hole is much likely to be an unstable solution of the AdS/BCFT holographic model.

Our analysis points to the fact that, from a charged black hole with tensionless RS brane, it cannot emerge a complete picture of a quantum Hall system. To find a more realistic AdS/BCFT description of the QHE one needs to construct solutions with non-vanishing tension of the RS brane. Such solutions should enable us to describe the configurations shown in figure~\ref{fig:gap}(d). We expect that the new solutions should have the following properties: namely, we expect that the dual system exhibits a gap in the ``bulk", an edge current running along its perimeter and that the tension of the RS branes should be somewhat related to the Fermi energy of the dual system. As a result the conductivities will be modified; in particular, the transverse electric conductivity in the ``bulk" and near the edges may differ, the ratio of the transverse heat and electric conductivities may satisfy a Wiedemann-Franz law~(\ref{WFlaw2}) modified by the interaction of the charge carriers.

\paragraph{Acknowledgements} We would like to thank E.~Akhmedov, A.~Dymarsky, A.~Ferraz, R.~Jackiw, H.~Johannesson, U.~Kol, G.~Semenoff and especially M.~Portnoi for useful discussions. DM is grateful to the Particle Physics Group at Tel Aviv University, where this work was initiated, for the warm hospitality. This work was supported by the Brazilian Ministry of Science and Technology, CAPES Foundation -- Brazil, and the IIP-MIT exchange program. The work of DM was also supported through the contract \#8206 with the Ministry of Education and Science of the Russian Federation, grant RFBR 11-02-01220 and grant of the President of the Russian Federation for Support of Scientific Schools NSh-3349.2012.2. The work of PS was also supported by the "Iniziativa Specifica INFN"  FI11 and by the EC Seventh Framework Programme (FP7/2007-2013) under grant \#295234-FP7-PEOPLE-2011-IRSES (project QICFT).



\begin{thebibliography}{99}

\bibitem{Takayanagi:2011zk}
  T.~Takayanagi,
  ``Holographic Dual of BCFT,''
  Phys.\ Rev.\ Lett.\  {\bf 107} (2011) 101602
  [arXiv:1105.5165 [hep-th]].

\bibitem{AdS/CFT}
  J.~M.~Maldacena,
  ``The Large N limit of superconformal field theories and supergravity,''
  Adv.\ Theor.\ Math.\ Phys.\  {\bf 2} (1998) 231
  [hep-th/9711200].

\bibitem{Ryu:2006bv}
  S.~Ryu and T.~Takayanagi,
  ``Holographic derivation of entanglement entropy from AdS/CFT,''
  Phys.\ Rev.\ Lett.\  {\bf 96} (2006) 181602
  [hep-th/0603001].

\bibitem{Randall:1999vf}
  L.~Randall and R.~Sundrum,
  ``An alternative to compactification,''
  Phys.\ Rev.\ Lett.\  {\bf 83} (1999) 4690
  [hep-th/9906064].

\bibitem{superconductivity}
 S.~S.~Gubser,
  ``Breaking an abelian gauge symmetry near a black hole horizon,''
  Phys.\ Rev.\ D {\bf 78} (2008) 065034
  [arXiv:0801.2977 [hep-th]].

  S.~A.~Hartnoll, C.~P.~Herzog and G.~T.~Horowitz,
  ``Building a Holographic Superconductor,''
  Phys.\ Rev.\ Lett.\  {\bf 101} (2008) 031601
  [arXiv:0803.3295 [hep-th]].

\bibitem{StrangeMetals}  H.~Liu, J.~McGreevy and D.~Vegh,
  ``Non-Fermi liquids from holography,''
  Phys.\ Rev.\ D {\bf 83} (2011) 065029
  [arXiv:0903.2477 [hep-th]].

\bibitem{Fujita:2012fp}
  M.~Fujita, M.~Kaminski and A.~Karch,
  ``SL(2,Z) Action on AdS/BCFT and Hall Conductivities,''
  JHEP {\bf 1207} (2012) 150
  [arXiv:1204.0012 [hep-th]].

\bibitem{Laughlin:1981jd}
  R.~B.~Laughlin,
  ``Quantized Hall conductivity in two-dimensions,''
  Phys.\ Rev.\ B {\bf 23} (1981) 5632.

\bibitem{FQHE} R.~B.~Laughlin,
  ``Anomalous quantum Hall effect: An Incompressible quantum fluid with fractionally charged excitations,''
  Phys.\ Rev.\ Lett.\  {\bf 50} (1983) 1395.

\bibitem{CompositeModel} J.~K.~Jain,
  ``Composite fermion approach for the fractional quantum Hall effect,''
  Phys.\ Rev.\ Lett.\  {\bf 63} (1989) 199.

  J.~K.~Jain,
  ``Theory of the fractional quantum Hall effect,''
  Phys.\ Rev.\ B {\bf 41} (1990) 7653.

\bibitem{Girvin}
  ``The Quantum Hall Effect, 2nd ed.," Eds. R.~E.~Prange and S.~M.~Girvin, Springer-Verlag, Berlin (1990).

  X-G.~Wen,
  ``Quantum Field Theory of Many-Body Systems,''
  Oxford University Press (2004).

\bibitem{Seso}
  R.~Jackiw,
  ``Fractional Charge and Zero Modes for Planar Systems in a Magnetic Field,''
  Phys.\ Rev.\ D {\bf 29} (1984) 2375
  [Erratum-ibid.\ D {\bf 33} (1986) 2500].

  J.~Avron, R.~Seiler and B.~Simon,
  ``Quantization of the Hall conductance for general multiparticle Schrodinger operators,"
  Phys.\ Rev.\ Lett.\  {\bf 54} (1985) 259.

  G.~W.~Semenoff and P.~Sodano,
  ``Nonabelian Adiabatic Phases And The Fractional Quantum Hall Effect,''
  Phys.\ Rev.\ Lett.\  {\bf 57} (1986) 1195.

  G.~W.~Semenoff, P.~Sodano and Y.~-S.~Wu,
  ``Renormalization Of The Statistics Parameter In Three-dimensional Electrodynamics,''
  Phys.\ Rev.\ Lett.\  {\bf 62} (1989) 715.

\bibitem{Polychronakos:1986me}
  A.~P.~Polychronakos,
  ``Topological mass quantization and parity violation in (2+1)-dimensional QED,''
  Nucl.\ Phys.\ B {\bf 281} (1987) 241.

  A.~P.~Polychronakos,
  ``On the quantization of the coefficient of the abelian Chern-Simons term,''
  Phys.\ Lett.\ B {\bf 241} (1990) 37.

  N.~Bralic, C.~D.~Fosco and F.~A.~Schaposnik,
  ``On the quantization of the abelian Chern-Simons coefficient at finite temperature,''
  Phys.\ Lett.\ B {\bf 383} (1996) 199
  [hep-th/9509110].

\bibitem{Kane} C.~I.~Kane, M.~P.~A.~Fisher,
  ``Quantized thermal transport in the fractional Quantum Hall Effect,''
  Phys.\ Rev.\ B {\bf 55} (1997) 15832
 [cond-mat/9603118].

\bibitem{Halperin:1981ug}
  B.~I.~Halperin,
  ``Quantized Hall conductance, current carrying edge states, and the existence of extended states in a two-dimensional disordered potential,''
  Phys.\ Rev.\ B {\bf 25} (1982) 2185.

\bibitem{Hartnoll:2007ai}
  S.~A.~Hartnoll and P.~Kovtun,
  ``Hall conductivity from dyonic black holes,''
  Phys.\ Rev.\ D {\bf 76} (2007) 066001
  [arXiv:0704.1160 [hep-th]].

\bibitem{Hartnoll:2007ih}
  S.~A.~Hartnoll, P.~K.~Kovtun, M.~Muller and S.~Sachdev,
  ``Theory of the Nernst effect near quantum phase transitions in condensed matter, and in dyonic black holes,''
  Phys.\ Rev.\ B {\bf 76} (2007) 144502
  [arXiv:0706.3215 [cond-mat.str-el]].

\bibitem{Davis:2008nv} J.~L.~Davis, P.~Kraus and A.~Shah,
  ``Gravity Dual of a Quantum Hall Plateau Transition,''
  JHEP {\bf 0811} (2008) 020
  [arXiv:0809.1876 [hep-th]].

M.~Fujita, W.~Li, S.~Ryu and T.~Takayanagi,
  ``Fractional Quantum Hall Effect via Holography: Chern-Simons, Edge States, and Hierarchy,''
  JHEP {\bf 0906} (2009) 066
  [arXiv:0901.0924 [hep-th]].

O.~Bergman, N.~Jokela, G.~Lifschytz and M.~Lippert,
  ``Quantum Hall Effect in a holographic model,''
  JHEP {\bf 1010} (2010) 063
  [arXiv:1003.4965 [hep-th]].

\bibitem{Nozaki:2012qd}
  M.~Nozaki, T.~Takayanagi and T.~Ugajin,
  ``Central Charges for BCFTs and Holography,''
  JHEP {\bf 1206} (2012) 066
  [arXiv:1205.1573 [hep-th]].

\bibitem{Fujita:2011fp}
  M.~Fujita, T.~Takayanagi and E.~Tonni,
  ``Aspects of AdS/BCFT,''
  JHEP {\bf 1111} (2011) 043
  [arXiv:1108.5152 [hep-th]].

\bibitem{Henningson:1998gx}
  M.~Henningson and K.~Skenderis,
  ``The holographic Weyl anomaly,''
  JHEP {\bf 9807} (1998) 023
  [hep-th/9806087].

\bibitem{Hayward:1993my}
  G.~Hayward,
  ``Gravitational action for space-times with nonsmooth boundaries,''
  Phys.\ Rev.\ D {\bf 47} (1993) 3275.

\bibitem{GKPW}
  S.~S.~Gubser, I.~R.~Klebanov and A.~M.~Polyakov,
  ``Gauge theory correlators from noncritical string theory,''
  Phys.\ Lett.\ B {\bf 428} (1998) 105
  [hep-th/9802109].

  E.~Witten,
  ``Anti-de Sitter space and holography,''
  Adv.\ Theor.\ Math.\ Phys.\  {\bf 2} (1998) 253
  [hep-th/9802150].

  D.~T.~Son and A.~O.~Starinets,
  ``Minkowski space correlators in AdS / CFT correspondence: recipe and applications,''
  JHEP {\bf 0209} (2002) 042
  [hep-th/0205051].

  C.~P.~Herzog and D.~T.~Son,
  ``Schwinger-Keldysh propagators from AdS/CFT correspondence,''
  JHEP {\bf 0303} (2003) 046
  [hep-th/0212072].

\bibitem{MagnetizationCurrents} N.~R.~Cooper, B.~I.~Halperin, I.~M.~Ruzin,
``Thermoelectric response of an interacting two-dimensional electron gas in a quantizing magnetic field,''
  Phys.\ Rev.\ B {\bf 55} (1997) 2344.

\bibitem{Faulkner:2009wj}
  T.~Faulkner, H.~Liu, J.~McGreevy and D.~Vegh,
  ``Emergent quantum criticality, Fermi surfaces, and AdS(2),''
  Phys.\ Rev.\ D {\bf 83} (2011) 125002
  [arXiv:0907.2694 [hep-th]].

\bibitem{Stripes} D.~B.~Chklovskii, B.~I.~Shklovskii, L.~I.~Glazman,
``Electrostatics of edge channels,''
  Phys.\ Rev.\ B {\bf 46} (1992) 4026.

  N.~B.~Zhitenev, R.~J.~Haug, K.~von~Klitzing, K.~Eberl,
  ``Time-resolved measurements of transport in edge channels,''
  Phys.\ Rev.\ Lett.\  {\bf 71} (1993) 2292.

  S.~W.~Hwang, D.~C.~Tsui, M.~Shayegan,
  ``Experimental evidence for finite-width edge channels in integer and fractional quantum Hall effects,''
  Phys.\ Rev.\ B {\bf 48} (1993) 8161.

\bibitem{UpstreamModes}  A.~H.~MacDonald,
  ``Edge states in the fractional-quantum-Hall-effect regime,''
  Phys.\ Rev.\ Lett.\  {\bf 64} (1990) 220.

X.~G.~Wen,
  ``Gapless boundary excitations in the quantum Hall states and in the chiral spin states,''
  Phys.\ Rev.\ B {\bf 43} (1991) 11025.

X.~G.~Wen,
  ``Electrodynamical properties of gapless edge excitations in the fractional quantum Hall states,''
  Phys.\ Rev.\ Lett.\  {\bf 64} (1990) 2206.

X.~G.~Wen,
  ``Edge transport properties of the fractional quantum Hall states and weak-impurity scattering of a one-dimensional charge-density wave,''
  Phys.\ Rev.\ B {\bf 44} (1991) 5708.

M.~D.~Johnson, A.~H.~MacDonald,
  ``Composite edges in the $\nu=2/3$ fractional quantum Hall effect,''
  Phys.\ Rev.\ Lett.\  {\bf 67} (1991) 2060.

\bibitem{Venkatachalam} V.~Venkatachalam, S.~Hart, L.~Pfeiffer, K.~West, A.~Yacoby,
 ``Local thermometry of neutral modes on the quantum hall edge,"
  Nat.\ Phys.\ {\bf 8} (2012), 676.

\end{thebibliography}
\end{document}